\documentclass[aps,prl,twocolumn,noshowpacs,a4paper,superscriptaddress]{revtex4-1}

\usepackage{graphicx}
\usepackage{color}
\usepackage{textcomp}
\usepackage{units}
\usepackage{amssymb,amsfonts,amsmath}
\usepackage{wasysym}
\usepackage[centerlast]{caption}

\renewcommand{\_}[1]{_\mathrm{#1}}
\newcommand{\bra}{\langle}
\newcommand{\ket}{\rangle}
\newcommand{\ev}[1]{\langle {#1} \rangle}

\begin{document}

\title{Real-time observation of fluctuations at the driven-dissipative Dicke phase transition}
\author{Ferdinand Brennecke}
\author{Rafael Mottl}\affiliation{Institute for Quantum Electronics, ETH Z\"{u}rich,
CH--8093 Z\"{u}rich, Switzerland}
\author{Kristian Baumann}\affiliation{Institute for Quantum Electronics, ETH Z\"{u}rich,
CH--8093 Z\"{u}rich, Switzerland}\affiliation{ present address:
Departments of Applied Physics, Physics and E.L. Ginzton Laboratory,
Stanford University, Stanford, California 94305, USA}
\author{Renate Landig}
\author{Tobias Donner} \email{Email: donner@phys.ethz.ch}
\author{Tilman Esslinger}\affiliation{Institute for Quantum Electronics, ETH Z\"{u}rich,
CH--8093 Z\"{u}rich, Switzerland}

\begin{abstract}
We experimentally study the influence of dissipation on the driven
Dicke quantum phase transition, realized by coupling external
degrees of freedom of a Bose-Einstein condensate to the light field
of a high-finesse optical cavity. The cavity provides a natural
dissipation channel, which gives rise to vacuum-induced fluctuations
and allows us to observe density fluctuations of the gas in
real-time. We monitor the divergence of these fluctuations over two
orders of magnitude while approaching the phase transition and
observe a behavior which significantly deviates from that expected
for a closed system. A correlation analysis of the fluctuations
reveals the diverging time scale of the atomic dynamics and allows
us to extract a damping rate for the external degree of freedom of
the atoms. We find good agreement with our theoretical model
including both dissipation via the cavity field and via the atomic
field. Utilizing a dissipation channel to non-destructively gain
information about a quantum many-body system provides a unique path
to study the physics of driven-dissipative systems.
\end{abstract}

\maketitle

Experimental progress in the creation, manipulation and probing of
atomic quantum gases has made it possible to study highly controlled
many-body systems and to access their phase transitions. This new
approach to quantum many-body physics has substantiated the notion
of quantum simulation for key models of condensed matter physics
\cite{Bloch2008,Lewenstein2012}. There has been increasing interest
in  generalizing such an approach to zero-temperature or quantum
phase transitions away from thermal equilibrium, as occurring in
driven-dissipative systems
\cite{Mitra2006,Diehl2008,DallaTorre2010,Torre2012b,Sieberer,Kessler2012}.
Amongst the most tantalizing questions is how vacuum fluctuations
from the environment influence the fluctuation spectrum at a phase
transition via quantum backaction. Coupling quantum gases to the
field of an optical cavity is a particularly promising approach to
realize a driven-dissipative quantum many-body system with a well
understood and controlled dissipation channel. A further advantage
of this scheme is that the dissipation channel of the cavity mode
can be directly used to investigate the system in a non-destructive
way via the leaking cavity field \cite{Mekhov2007a}.

\begin{figure}[ht!]
\includegraphics[width = 80mm]{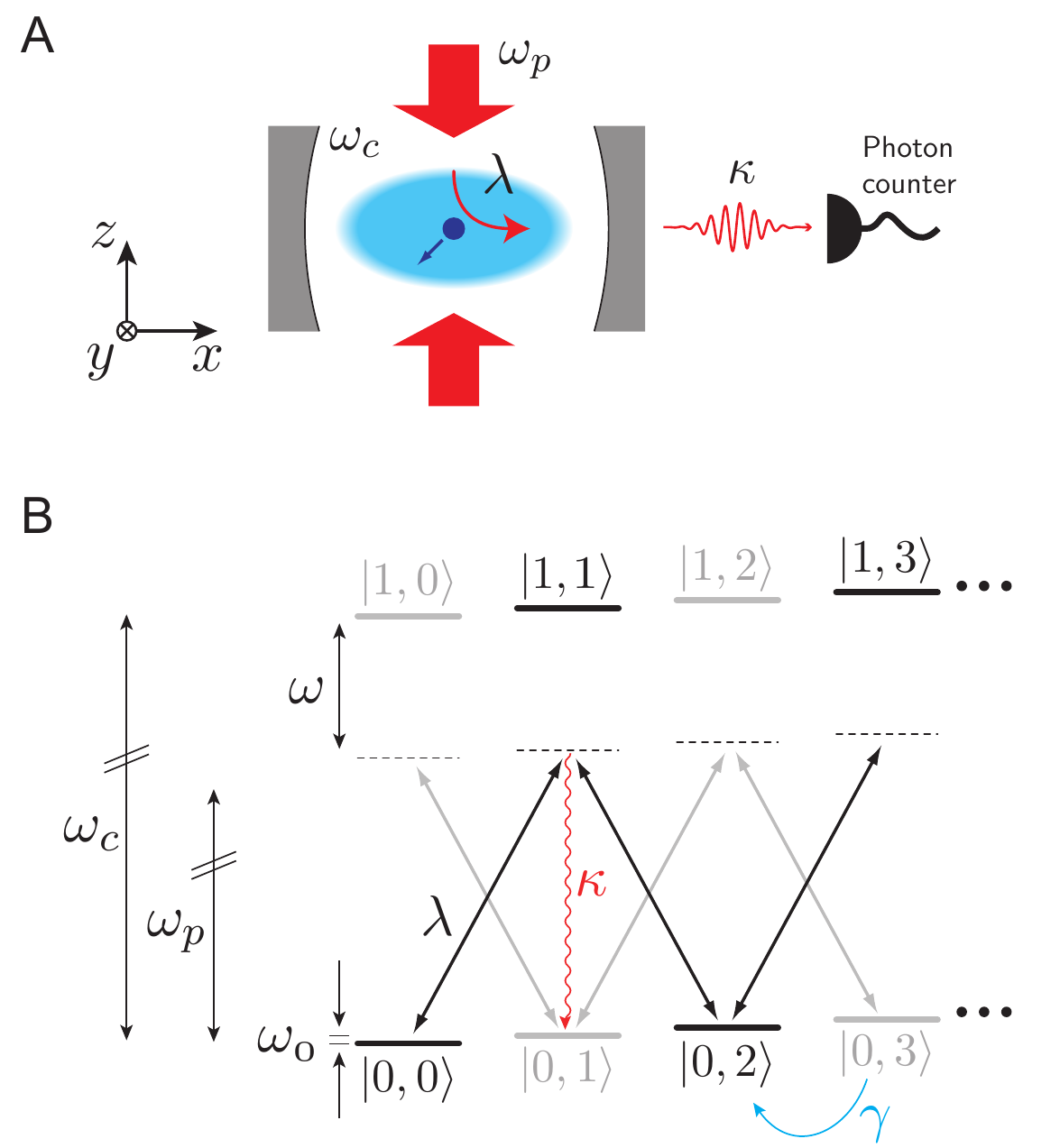}
\caption{\textbf{A.} Experimental scheme: A transverse pump field
(red) couples an excited momentum mode of a BEC (blue) to a cavity
mode via collective light scattering at rate $\lambda$. The cavity
provides a loss channel for the system through which photons can
escape. Density fluctuations are inferred from the detected cavity
output field. \textbf{B.} Level scheme of the system after
elimination of the electronically excited atomic states. The ground
state of the closed coupled system is given by a coherent
superposition of states $|n_a,n_b\rangle$ with even parity (black
symbols). Here, $n_a$ is the number of intracavity photons and $n_b$
is the number of momentum excitations. Decay processes drive the
system to a steady state which also includes an incoherent
population of states with odd parity (grey symbols), caused by
either the decay of a cavity photon at rate $2\kappa$ or the decay
of a momentum excitation at rate $2\gamma$. The depicted level
scheme is restricted to $n_a \leq 1$. }\label{figure1}
\end{figure}

Combining the experimental setting of cavity quantum electrodynamics
with that of quantum gases \cite{Brennecke2007, Colombe2007,
Gupta2007, Wolke2012, Ritsch2012} led to the observation of quantum
backaction heating caused by cavity dissipation \cite{Murch2008,
Brahms2012}, as well as to the realization of the non-equilibrium
Dicke quantum phase transition \cite{Baumann2010}. Here, we study
the influence of cavity dissipation on the fluctuation spectrum at
the Dicke phase transition by connecting these approaches. We
non-destructively observe diverging fluctuations of the order
parameter when approaching the critical point, and find a distinct
difference with respect to predictions for the closed, i.e.
non-dissipative system.

In our experimental system, density wave excitations in a
Bose-Einstein condensate (BEC) are coupled via a coherent laser
field to the mode of a standing-wave optical cavity. For strong
enough coupling this causes a spatial self-organization of the atoms
on a wavelength-periodic checkerboard pattern which is a realization
of the driven-dissipative Dicke phase transition \cite{Baumann2010,
Nagy2010a}. The fluctuations triggering the phase transition are
atomic density fluctuations. They are generated by long-range
atom-atom interactions which are mediated by exchange of cavity
photons \cite{Mottl2012a}. In the presence of cavity decay, vacuum
fluctuations enter the cavity, interfere with the coherent pump
laser field, and drive the system to a steady state with increased
density fluctuations \cite{Nagy2011,Oztop2012}. In turn, the cavity
decay offers natural access to the system properties via the light
field leaking out of the cavity, which allows us to measure the
density fluctuations in real-time \cite{Mekhov2007a}. Except for the
natural quantum backaction of this continuous measurement process
\cite{Braginsky1993}, the system remains unperturbed by our
observation.

\section{System description}
\subsection{Hamiltonian dynamics} As described in our previous
work \cite{Baumann2010, Mottl2012a}, we place a BEC of $N$ atoms
inside an ultrahigh-finesse optical cavity and pump the atoms
transversally with a far-detuned standing-wave laser field
(Fig.~\ref{figure1}A). The closed-system dynamics is described by
the Dicke model (\cite{Baumann2010,Nagy2010a} and SI),
\begin{equation}\label{Dicke}
    \hat{H} = \hbar \omega \hat{a}^{\dag} \hat{a}+ \hbar \omega_0 \hat{J}_z + \frac{2 \hbar \lambda}{\sqrt{N}}
    (\hat{a}+\hat{a}^{\dag}) \hat{J}_x \,,
\end{equation}
where $\omega$ denotes the detuning between pump laser frequency
$\omega_\textrm{p}$ and dispersively shifted cavity resonance
frequency $\omega_\textrm{c}$, and $\hbar$ is Planck's constant
divided by $2\pi$. The annihilation operator of a cavity photon in a
frame rotating at $\omega_{\textrm{p}}$ is given by $\hat{a}$. The
atomic dynamics is captured in a two-mode description, consisting of
the macroscopically populated zero-momentum mode $\psi_0$ of the
BEC, and an excited momentum mode $\psi_1$, carrying in a symmetric
superposition one photon momentum along the $\pm x$ direction and
one along the $\pm z$ direction. This defines an effective two-level
system with energy splitting $\hbar \omega_0 = \hbar^2 k^2/m$, where
$k$ denotes the optical wavevector and $m$ the atomic mass. The
atomic ensemble of $N$ such two-level systems can be described by
collective spin operators $\hat{J}_x$, $\hat{J}_y$ and $\hat{J}_z$.
The expectation value $\langle \hat{J}_x \rangle$ measures the
checkerboard density modulation which results from the interference
between coherent populations of the two matter wave modes and can be
identified as order parameter of the phase transition. The coupling
strength $\lambda \propto \sqrt{P}$ between atomic motion and light
field can be experimentally controlled via the power $P$ of the
transverse pump field, and represents the collective two-photon Rabi
frequency of the underlying scattering process between pump and
cavity field (Fig.~\ref{figure1}A). When $\lambda$ reaches the
critical coupling strength $\lambda_{\textrm{cr}}=\frac{\sqrt{\omega
\omega_0}}{2}$, Hamiltonian~\eqref{Dicke} gives rise to a
second-order quantum phase transition \cite{Emary2003a} towards a
phase characterized by a non-zero order parameter $\langle \hat{J}_x
\rangle \neq 0$, and a coherent cavity field, $\langle
\hat{a}\rangle \neq 0$. Below the critical point, the system is in
the normal phase, $\langle \hat{J}_x \rangle = \langle
\hat{a}\rangle = 0$, where only fluctuations of the order parameter,
$\langle \hat{J}^2_x \rangle \neq 0$, give rise to an incoherent
cavity field with $\langle \hat{a}^{\dag}\hat{a}\rangle \neq 0$.

In the thermodynamic limit, the fluctuations of the order parameter
in the normal phase can be described with bosonic creation and
annihilation operators $\hat{b}^{\dag}$ and $\hat{b}$ according to
$J_x=\sqrt{N}(\hat{b}+\hat{b}^{\dag})/2$ (SI). The interaction term
in Eq.~\eqref{Dicke} then becomes $\hbar \lambda
(\hat{a}+\hat{a}^{\dag})(\hat{b}+\hat{b}^{\dag})$, and couples the
bare states $|n_a,n_b\rangle$ under parity conservation of the total
number of excitations $n_a+n_b$. Here, $n_a$ is the number of
photons stored in the cavity and $n_b$ is the number of excitations
in the momentum mode $\psi_1$ (Fig.~\ref{figure1}). The ground state
of the closed, coupled system is a two-mode squeezed state
\cite{Emary2003, Milburn2008} with admixtures of the even parity
states only ($|0,0\rangle, |1,1\rangle, |0,2\rangle,\ldots$). For
$\omega \gg \omega_0$, the cavity is almost only virtually
populated, i.e. the admixture of states with $n_a\neq0$ is
suppressed by $\omega_0/\omega$. The quantum fluctuations of the
Hamiltonian system then correspond dominantly to pairs of atoms in
the excited momentum mode $\psi_1$. They are created and annihilated
by quasi-resonant scattering of a pump photon into the cavity mode
and back into the pump field at a rate $\frac{\lambda^2}{\omega}$.
Towards $\lambda_{\textrm{cr}}$, the variance $\langle
(\hat{b}+\hat{b}^{\dag})^2\rangle$ of the resulting density
fluctuations diverges, while the gas still does not show a density
modulation ($\langle \hat{b}+\hat{b}^{\dag}\rangle = 0$).

\subsection{Dissipative dynamics} In the case of the open system,
the tiny population of states with $n_a \ne 0$ becomes important. As
this decays via cavity dissipation, the ladder of states with odd
parity ($|1,0\rangle, |0,1\rangle, |1,2\rangle, |0,3\rangle,\ldots$)
is incoherently populated (Fig.~\ref{figure1}). The microscopic
process corresponds to the loss of a cavity photon at one of the
mirrors before the coherent scattering back into the pump beam can
be completed. The system will thus leave its ground state and
irreversibly evolve into a non-equilibrium steady state with
additional density fluctuations and a constant energy flow from the
pump laser to the cavity output. The variance of the resulting
incoherent cavity population $\langle \hat{a}^{\dag} \hat{a}
\rangle$ has been predicted to diverge at $\lambda_\textrm{cr}$ with
a critical exponent of 1.0 compared to the closed system exponent of
0.5 \cite{Nagy2011,Oztop2012}. The depletion of the ground state
takes place at rate $\kappa_{\textrm{eff}}=
\frac{\lambda^2}{\omega^2 + \kappa^2}\cdot\kappa $, where $\kappa$
is the decay rate of the cavity field \cite{Nagy2010a}. In the
experiment, $\kappa_{\textrm{eff}}$ can be tuned and we choose
$\omega \approx 8 \kappa$, such that the rate of decay processes is
almost an order of magnitude below the long-range interaction rate
$\lambda^2/\omega$.

The observable in our experiment is light leaking out of the cavity.
Since for $\kappa \gg \omega_0$ the cavity field adiabatically
follows the atomic motion, the cavity output field provides a
sensitive tool to monitor the order parameter and its fluctuations
in real time (SI, \cite{Mekhov2007a}). Checkerboard density
fluctuations with variance $\langle
(\hat{b}+\hat{b}^{\dag})^2\rangle$ induce a finite incoherent
population of the cavity field according to $\langle \hat{a}^{\dag}
\hat{a} \rangle  = \frac{1}{4}
\frac{\omega_0}{\omega}\left(\frac{\lambda}{\lambda_{\textrm{cr}}}\right)^2
\langle (\hat{b}+\hat{b}^{\dag})^2\rangle$. Cavity decay amounts to
a continuous measurement of the intracavity light field and causes,
due to inherent matter-light entanglement, a quantum backaction upon
the atomic system \cite{Braginsky1993}. The role of the photons
leaking out of the cavity is thus two-fold: they drive the system to
a steady state of enhanced fluctuations and reveal real-time
information about the total density fluctuations.

\section{Results}
\subsection{Data acquisition} 
\begin{figure}[ht!]
\centering
\includegraphics[width = 80mm]{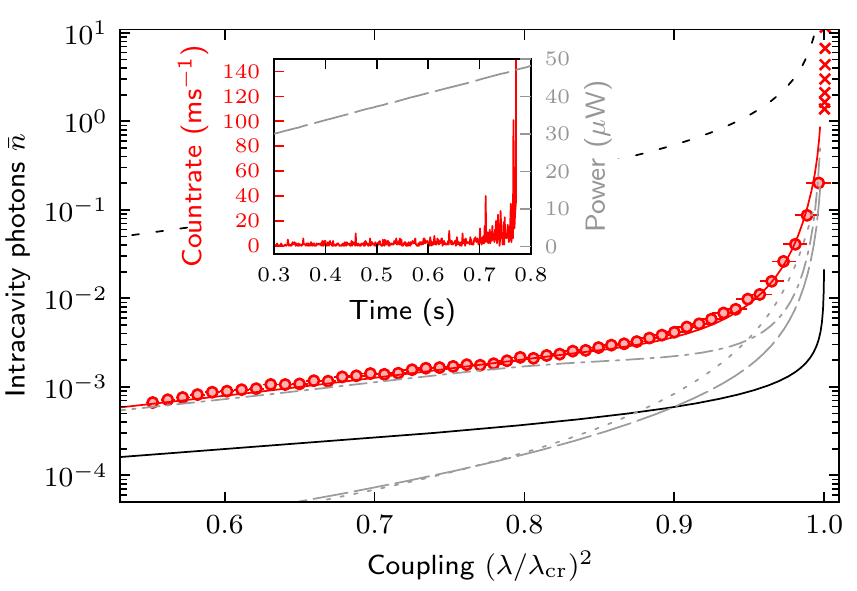}
\caption{Mean intracavity photon number $\bar{n}$ (red symbols) as a
function of coupling. Circles (crosses) indicate data in the normal
(ordered) phase, the errorbars display the statistical error. The
calculated expectations for the closed system are shown as solid
black line. Our open-system description (solid red line) includes
cavity field fluctuations due to the decay of photons and momentum
excitations (grey dashed-dotted line) and due to the finite
temperature of the BEC (grey dashed line), as well as a symmetry
breaking coherent cavity field (grey dotted line). We also show the
calculated fluctuations if the atomic damping rate $\gamma$ would
vanish (black dashed line). \textbf{Inset:} The raw data of a single
run (red line) is displayed together with the measured transverse
pump power (grey dashed line) as a function of time. The sudden
increase in the photon countrate clearly marks the transition
point.}\label{figure2}
\end{figure}

Using this concept, we experimentally
observe density fluctuations of the atomic ensemble in the normal
phase while approaching the phase transition. We prepare the system
with $N=1.6(2)\cdot 10^5$ $^{87}$Rb atoms at an intermediate
coupling of $(\lambda/\lambda_{\textrm{cr}})^2 \approx 0.55 $ and at
a detuning of $\omega=\unit[2\pi \cdot 10.0(5)]{MHz}$. Then, the
transverse pump-laser power is linearly increased within a data
acquisition time of 0.8\,s to a value slightly beyond the critical
point. For our parameters, $\omega_0=2\pi\cdot\unit[8.3(2)]{kHz}$
and $\kappa=2\pi\cdot\unit[1.25(5)]{MHz}$ \cite{Mottl2012a}, the
rate $\kappa_{\textrm{eff}}$ at which the steady state is approached
is $2\pi \cdot \unit[1]{kHz}$ for $\lambda=\lambda_{\textrm{cr}}$
\cite{Nagy2010a}. We can therefore assume the system to be in steady
state throughout the measurement. The inset in Fig.~\ref{figure2}
displays the data of a single experimental run, where we monitor the
stream of photons leaking out of the cavity with a single-photon
counting module. From the photon count rate $r$ we deduce the
intracavity photon number $\bar{n}=(r-r_\textrm{b})/2\kappa\eta$,
taking into account the measured total detection efficiency of
$\eta=$\unit[5(1)]{\%} and the independently calibrated background
count rate $r_\textrm{b}$. We observe a progressively increasing
photon count rate with increasing transverse pump laser power, until
a steep rise marks the transition point to the ordered phase. The
exact position of the transition depends on the total number of
atoms which fluctuates by \unit[10]{\%} between repeated
experimental runs. Therefore, we define a threshold for the count
rate to detect the transition point (SI). This allows us to convert
the time axis into linearly increasing coupling.

\subsection{Mean intracavity photon number} We average the signal
of 372 experimental runs and observe the divergence of the
intracavity photon number $\bar{n}$ over three orders of magnitude,
ending in a steep increase after passing the critical point
(Fig.~\ref{figure2}). We compared the measured intracavity photon
number with the cavity field fluctuations expected from the ground
state of the closed system \cite{Emary2003}. Our data clearly shows
an enhanced cavity field occupation with respect to the Hamiltonian
system (solid black line in Fig.~\ref{figure2}). This is in
accordance with the presented picture that cavity decay increases
fluctuations. Yet, the magnitude of the observed fluctuations is
well below the theoretical expectation \cite{Nagy2011,Oztop2012} for
a cavity decay at rate $\kappa$ (dashed black line in
Fig.~\ref{figure2}). This indicates the presence of a further
dissipation channel which damps out atomic momentum excitations.

\subsection{Correlation analysis} 
\begin{figure}
\centering
\includegraphics[width = 80mm]{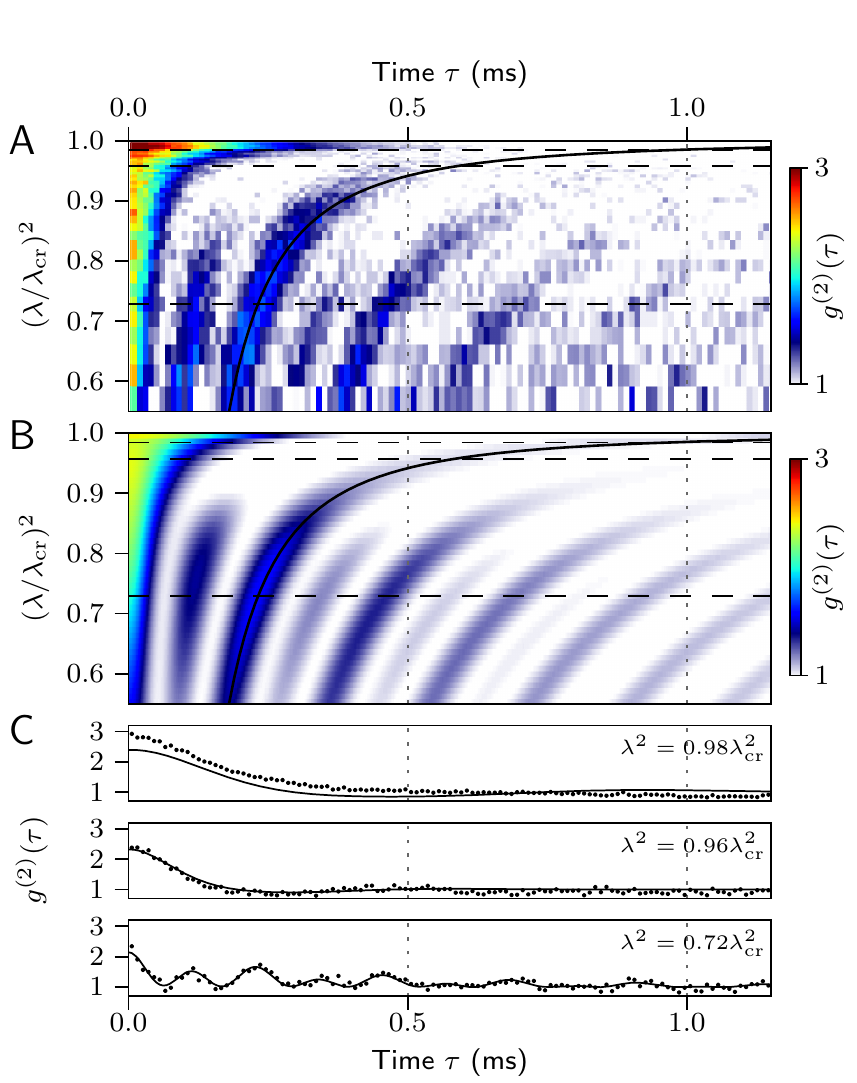}
\caption{Temporal correlations of the cavity output. Panel
\textbf{A} shows a color plot of the measured second-order
correlation function $g^{2}(\tau)$ as a function of time $\tau$ and
coupling $(\lambda/\lambda_{\textrm{cr}})^2$. The correlation time
increases with increasing coupling in agreement with the timescale
related to the lowest excitation energy $\omega_s$ of the coupled
system (solid black line). Panel \textbf{B} shows the correlations
$g^{2}(\tau)$ calculated from the full theoretical model with
parameters $\zeta$ and $\gamma$ adjusted to match the data. The
horizontal dashed lines indicate values of
$(\lambda/\lambda_{\textrm{cr}})^2$ along which the data is shown
(points) on the lower three panels \textbf{C} together with the full
theory.}\label{figure3}
\end{figure}

Additional insight into the fluctuation
dynamics and possible dissipative processes can be gained from a
correlation analysis of the cavity output field. We calculate the
second-order correlation function for all experimental data
contributing to Fig.~\ref{figure2}. Since the cavity field
adiabatically follows the atomic dynamics, its second-order
correlation function $g^{(2)}(\tau)\propto \langle
\hat{a}^{\dag}(\tau) \hat{a}^{\dag}(0) \hat{a}(0)
\hat{a}(\tau)\rangle$ is linked to the temporal correlation function
of the order parameter fluctuations $\langle \hat{J}_x^2(\tau)
\hat{J}_x^2(0) \rangle$. The evaluated correlations as a function of
time and coupling are shown in Fig.~\ref{figure3}, together with
cuts for specific coupling values. In contrast to a purely coherent
cavity output field, which would yield a flat correlation function,
we observe enhanced correlations for short times, followed by damped
oscillations. The frequency of these oscillations agrees with the
excitation energy of the coupled system, $\hbar \omega_{\textrm{s}}
= \hbar \omega_{\textrm{0}}
\sqrt{1-(\lambda/\lambda_{\textrm{cr}})^2}$, which softens with
increasing coupling and tends towards zero at the critical point.
This shows that the cavity output field indeed carries information
about the incoherent fluctuations of the system, and is consistent
with our previous measurement of a mode softening and a diverging
response \cite{Mottl2012a}. A vanishing excitation frequency
corresponds to a critical slowing down of the dynamics. Within our
measurement resolution, however, the system adiabatically follows
the steady state since the rate of change
$d/dt({\lambda}/\lambda_{\textrm{cr}})^2$ is only a few Hz
\cite{Baumann2011}.

We attribute the damping of the oscillations in $g^{(2)}(\tau)$ to
the decay of atomic momentum excitations. This constitutes an
additional dissipation channel caused by collisional and possibly
cavity-mediated coupling of momentum excitations to Bogoliubov modes
of the BEC \cite{Graham2000, Katz2002}. The observed decay rate of
$g^{(2)}(\tau)$ cannot be explained by a finite admixture of the
cavity field in the steady state, since $\omega$ exceeds $\omega_0$
by orders of magnitude in our system \cite{Nagy2011}.

The oscillations in the second-order correlation function exhibit an
overperiod which becomes more pronounced towards the critical point.
This indicates the presence of a finite coherent cavity field
amplitude $\alpha = \langle \hat{a} \rangle$ which we attribute to
the finite cloud size of the BEC and residual scattering of pump
light at the edges of the cavity mirrors \cite{Baumann2010}.
Interference between the coherent and incoherent cavity field
components then causes the observed overperiod in the correlation
function.

\subsection{Quantum Langevin description} To quantitatively describe
our observations, we developed a theoretical model based on coupled
quantum Langevin equations \cite{Dimer2007} capturing the dynamics
of the driven-dissipative system (SI). Our model explicitly takes
into account the dissipation of the cavity field at rate $\kappa$,
and furthermore a dissipation channel for excitations in the atomic
momentum mode $\psi_1$. For simplicity, this dissipation channel is
phenomenologically modeled by a thermal Markovian bath at the BEC
temperature of $\unit[100(20)]{nK}$ into which excitations in the
momentum mode $\psi_1$ decay at a rate $\gamma$ (SI). Due to the
softening excitation frequency $\omega_s$, the decay rate $\gamma$
is taken as a function of the coupling rate $\lambda$. Our model
further includes a small symmetry breaking field, which results in a
coherent cavity field amplitude $\alpha$ already below the critical
point. This is taken into account by renormalizing the order
parameter with a constant offset $\zeta$ in Eq.~\eqref{Dicke}
\cite{Baumann2010}.

From the solution of the quantum Langevin equations in the
thermodynamic limit we obtain the second-order correlation function
of the intracavity field in the steady state (SI). The free
parameters of our model description ($\zeta$ and $\gamma$) are
extracted from fits of the model to the correlation data
(Fig.~\ref{figure3}). We obtain an order parameter offset $\zeta =
60(7)$ at $\lambda=0$, which corresponds to 0.8$\permil$ of the
maximal possible order parameter $N/2$. This agrees with our earlier
investigation of the symmetry breaking field \cite{Baumann2011}.

\begin{figure}
\centering
\includegraphics[width = 80mm]{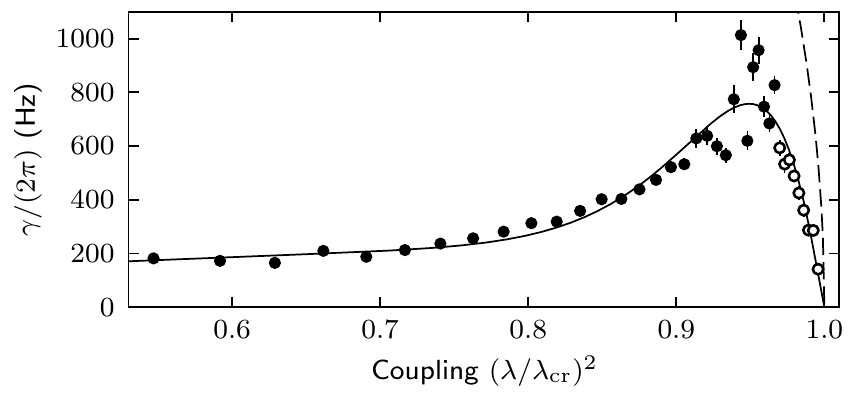}
\caption{Damping rate $\gamma$ (symbols) as a function of coupling
$(\lambda/\lambda_{\textrm{cr}})^2$ deduced from the cavity output
correlation data. Errorbars indicate the statistical error derived
from the fit. Open circles indicate the region above
$(\lambda/\lambda_{\textrm{cr}})^2=0.97$, where our theoretical
model deviates significantly from the data and values for $\gamma$
might be inaccurate. The solid line shows the phenomenological
function used to model the data in Fig.~\ref{figure2}A. The dashed
line shows the vanishing excitation frequency
$\omega_\textrm{s}/2\pi$ of the system.}\label{figure4}
\end{figure}

The extracted damping rate $\gamma$ is displayed in
Fig.~\ref{figure4} as a function of coupling. It increases with
increasing coupling, until it exhibits a cusp around 95\% of the
critical coupling and vanishes towards the critical point. We
attribute this behavior mainly to the softening of the excitation
frequency $\omega_\mathrm{s}$ (dashed line in Fig.~\ref{figure4},
\cite{Mottl2012a}) which influences the density of states into which
the momentum excitations in mode $\psi_1$ can decay. At the critical
point, this is expected to lead to the absence of damping of the
excited momentum mode \cite{Graham2000,PrivateComm}.

Our model describes our data very well for coupling values up to
$(\lambda/\lambda_{\textrm{cr}})^2 \simeq 0.97$. Above this value,
we observe enhanced correlations for small $\tau$ which are not
captured by the model, as can be seen in the uppermost subpanel of
Fig.~\ref{figure3}C. We believe that in this region technical
fluctuations, the dynamical change in the dispersive cavity shift
\cite{Keeling2010a}, finite-$N$ effects \cite{Konya2011}, and
population of higher order momentum states start to play a role.

Using the extracted atomic damping rate and symmetry breaking field
magnitude, we find very good agreement between the observed
intracavity photon number and our model (Fig.~\ref{figure2}). The
inclusion of atomic damping is crucial for the quantitative
description. While cavity decay is expected to lead to a strong
increase of the density fluctuations, atomic dissipation dominantly
damps out these momentum excitations, such that the total
fluctuations in the steady state are only moderately enhanced with
respect to the ground state fluctuations. Except for a small region
close to the critical point, the dominant contribution to the
observed fluctuations originates from vacuum input noise associated
with dissipation via the cavity (grey dashed-dotted line in
Fig.~\ref{figure2}). Only close to the critical point, fluctuations
from the thermal atomic bath are predicted to contribute because the
energy of the relevant mechanical excitation vanishes towards the
phase transition \cite{Mottl2012a}.

\subsection{Density fluctuations} 
\begin{figure}[ht!]
\centering
\includegraphics[width = 80mm]{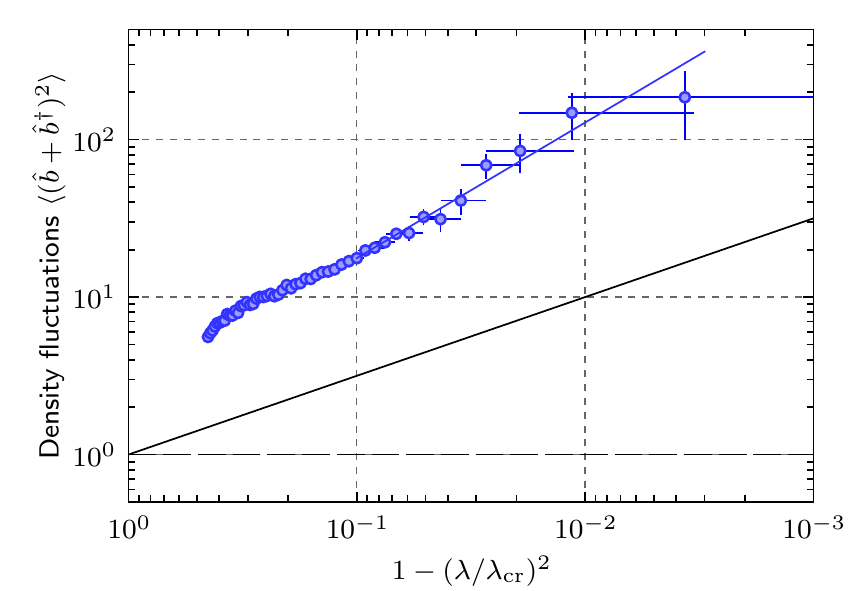}
\caption{Variance of the checkerboard density fluctuations of the
BEC, deduced from the intracavity photon number after subtracting
the coherent contribution. For comparison, we show the theory of the
closed system (black line) which diverges with a critical exponent
of $0.5$, and a linear fit (blue line) to the data for
$(\lambda/\lambda_{\textrm{cr}})^2\ge0.9$ that results in an
exponent of $0.9 \pm 0.1$. We also plot the expected fluctuations
for a BEC without coupling to the cavity field (black dashed line).
The horizontal errorbars indicate the statistical error, while the
vertical errorbars result from the uncertainty in the subtracted
coherent field component (SI).}\label{figure5}
\end{figure}

We infer the variance of the
density fluctuations $\langle(\hat{b}+\hat{b}^{\dag})^2 \rangle$ in
the normal phase by rescaling the intracavity photon number
$\bar{n}$ after subtracting the coherent fraction $|\alpha|^2$,
which was deduced from the correlation analysis
(Fig.~\ref{figure5}). Due to the uncertainties in the symmetry
breaking field $\zeta$, this procedure results in systematical
uncertainties of the deduced density fluctuations which are
reflected in the presented error bars. Our data, displayed on a
log-log scale, deviates clearly in both magnitude and scaling from
the expectations for the closed system. A linear fit (blue line) to
the data results in an exponent of 0.9($\pm$ 0.1). Scaling with
exponent 1.0 was predicted from open-system calculations in which
only cavity dissipation is taken into account \cite{Nagy2011,
Oztop2012}. The influence of the additional atomic dissipation rate
$\gamma$ on the scaling of the atomic density fluctuations depends
on the precise scaling of this damping rate when approaching the
critical point, which goes beyond the scope of this publication.

\section{Discussion}
From a more general perspective, driven systems, coupled via a
dissipation channel to a zero-temperature Markovian bath, are
expected to resemble classical critical behavior and can then be
characterized in steady state by an effective temperature which
depends on the considered observable \cite{Mitra2006, Diehl2008,
DallaTorre2010, Torre2012b, Gopalakrishnan2009}. In our system, the
zero-temperature bath is provided by the optical vacuum modes
outside the cavity. Verifying the fluctuation-dissipation theorem
for the order parameter in our system would allow to determine its
effective temperature. The theoretical expectation of a critical
exponent of 1.0 \cite{Nagy2011, Oztop2012} is a further indication
that systems undergoing a driven-dissipative phase transition can be
described to be effectively thermalized. However, answering the
question whether cavity dissipation completely destroys the quantum
character of the system, e.g. the entanglement between atomic and
light fields, remains a challenge for future experiments
\cite{Nagy2011, Dimer2007}.

\section{Conclusion and Outlook}
We have demonstrated the direct observation of diverging density
fluctuations in a quantum gas undergoing the driven-dissipative
Dicke phase transition. This experiment opens a route to study
quantum phase transitions in open systems under well controlled
conditions. Our method directly uses the cavity dissipation channel
to obtain real-time information on the fluctuations of the order
parameter. In a similar way, intriguing quantum many-body states
with long-range atom-atom interactions and the influence of
dissipation on them can be investigated e.g. by using multi-mode
cavities which allow to realize glassy and frustrated states of
matter \cite{Gopalakrishnan2009, Strack2011a}. Adding classical
optical lattices to the system would let the energy scale of contact
interactions enter the dynamics and should allow the exploration of
rich phase diagrams \cite{Larson2008a, Habibian2013}.

\section{Acknowledgements}

We acknowledge insightful discussions with I. Carusotto, S. Diehl,
P. Domokos, S. Gopalakrishnan, S. Huber, A. Imamoglu, M.
Paternostro, C. Rama, H. Ritsch, G. Szirmai, and H. T\"ureci.
Supported by Synthetic Quantum Many-Body Systems (European Research
Council advanced grant), Nanodesigning of Atomic and Molecular
Quantum Matter (European Union, Future and Emerging Technologies
open), National Centre of Competence in Research/Quantum Science and
Technology, and the European Science Foundation (POLATOM).

\appendix
\begin{widetext}

\begin{center}
\Large{\textbf{Supplementary Information}}
\end{center}

\section{Experimental details and data analysis}

The experimental sequence for the measurements presented is as
follows. After preparing an almost pure Bose-Einstein condensate in
a crossed-beam dipole trap which is centered with respect to the
TEM$_{00}$ cavity mode, the transverse pump power $P$ is increased
over $\unit[50]{ms}$ to a relative coupling strength of
$\lambda^2/\lambda^2_\mathrm{cr} \approx 0.55$. Subsequently, the
power $P$ is linearly increased over $\unit[0.8]{s}$ to slightly
beyond the critical pump power, while the stream of photons leaking
out of the cavity is recorded on a single-photon counting module.
Intracavity photons are detected with an efficiency of $5\%$,
limited mainly by the losses in the cavity mirrors and the detection
efficiency of the single-photon counter. We carefully calibrated the
background count rate of the photon detector in the absence of atoms
yielding a maximum rate of $\unit[341(62)]{/s}$. This results from
dark counts (at a rate of $100(10)/s$) and stray light of the
transverse pump beam and the dipole trap beams. The background
signal is subtracted from the overall count rate recorded in the
presence of atoms to obtain the mean intracavity photon number, as
shown in Figure 2.

A steep increase of the photon count rate indicates the phase
transition point and allows to convert the recording time axis into
a relative coupling axis $\lambda^2/\lambda_\mathrm{cr}^2$. To this
end we identify the timebin of $\unit[100]{\mu s}$ length in which
the photon count rate exceeds for the first time the value of
$\unit[18]{\mu s^{-1}}$ corresponding to a mean intracavity photon
number of approximately 49. According to the typical risetime of the
intracavity signal, the critical point is defined as the time
$\unit[1]{ms}$ prior to this timebin. The relative error of
$\lambda_\mathrm{cr}$ according to this procedure is given by
$5\cdot 10^{-4}$. Residual atom loss of $10\%$ during probing is
taken into account by rescaling the relative coupling axis according
to the proportionality $P_\mathrm{cr} \propto N^{-1}$. This is
justified by the resulting match between the oscillation frequency
of the second-order correlation function and the softening
excitation frequency, which was measured independently in
\cite{SMottl2012}.

For the correlation analysis, the recorded photon data is cut into
subtraces whose length decreases to a minimal value of
$\unit[4]{ms}$ as the critical point is approached. For each
subtrace the photon-photon correlation function is computed (using
time bins of $\unit[2]{\mu s}$). The normalized correlation
functions are finally averaged over all 372 experimental runs.

The temperature of the initially prepared Bose-Einstein condensate
was determined from absorption images to $T = \unit[65(20)]{nK}$.
During probing, this value is expected to increase by
$\unit[35]{nK}$ due to spontaneous emission and decay of atomic
momentum excitations into the bath of Bogoliubov modes (see below).
Further experimental details can be found in
\cite{SMottl2012,SBaumann2010}.

\section{Theoretical model}
\subsection{Two-mode description and Dicke model}

We model the dynamics of the transversally driven BEC-cavity system
in a two-mode description which formally is equivalent to the Dicke
Hamiltonian, as was shown previously \cite{SBaumann2010,SNagy2010}.
The relevant motional states which are coupled via two-photon
transitions between pump and cavity fields are given by the
macroscopically populated condensate mode $\psi_0$ and the
motionally excited mode $\psi_1$ which carries in a coherent
superposition momenta $(p_x, p_z) = (\pm \hbar k, \pm\hbar k)$ along
the cavity ($x$) and pump ($z$) directions. To first order, these
matter wave modes are separated in energy by $\hbar \omega_0 = 2
\hbar \omega_R$, with recoil frequency $\omega_R = \hbar k^2/2m$,
optical wavevector $k = 2\pi/\lambda$ and atomic mass $m$. Atomic
s-wave scattering and a weak $\lambda/2$-periodic lattice potential
caused by the transverse pump field induce a small shift of
$\hbar\omega_0$ \cite{SMottl2012}.

After expanding the atomic field operator in this two-mode basis,
$\hat{\Psi} = \hat{c}_0 \psi_0 + \hat{c}_1 \psi_1$, and introducing
corresponding collective spin-$N/2$ operators (where $N$ denotes the
total number of atoms)
\begin{equation}
\hat{\bf{J}} = \left(
                 \begin{array}{c}
                   \hat{J}_x\\
                   \hat{J}_y\\
                   \hat{J}_z\\
                 \end{array}
               \right)
 = \left(
     \begin{array}{c}
       \frac{1}{2}(\hat{c}_1^\dag\hat{c}_0+\hat{c}_0^\dag\hat{c}_1)
       \\
       \frac{1}{2i}(\hat{c}_1^\dag\hat{c}_0-\hat{c}_0^\dag\hat{c}_1) \\
       \frac{1}{2}(\hat{c}_1^\dag\hat{c}_1-\hat{c}_0^\dag\hat{c}_0) \\
     \end{array}
   \right),
\end{equation}
the many-body Hamiltonian of the BEC-cavity system is given by the
Dicke Hamiltonian
\begin{equation}
\label{eq:Dicke Hamiltonian} \hat{H}/\hbar =  \omega \hat{a}^\dag
\hat{a} + \omega_0 \hat{J}_z + 2\lambda /\sqrt{N} (\hat{a}+
\hat{a}^\dag) \hat{J}_x\,.
\end{equation}
Note that the time-dependence of the driven system was formally
eliminated by moving into a reference frame rotating at the pump
laser frequency $\omega_p$. In equation \eqref{eq:Dicke
Hamiltonian}, $\hat{a}$ annihilates a photon in the cavity mode with
resonance frequency $\omega_c$, which in the experiment is
blue-detuned from the pump laser frequency $\omega_p$ by the amount
$\omega = \omega_c - \omega_p = 2\pi \times \unit[10]{MHz}$. Here,
the dispersive cavity shift caused by the bare condensate ($2\pi
\times \unit[4]{MHz}$ for $N = 1.6 \times 10^5$ atoms) is already
included in $\omega_c$. In the non-organized phase ($\lambda <
\lambda_\mathrm{cr}$), the change of the dispersive cavity shift due
to the small population of the motional excited state $\psi_1$ can
be neglected for our experimental parameters (see however
\cite{SKeeling2010,SBhaseen2012}). Working at large pump-cavity
detuning, we also neglect the substructure of the atoms-shifted
cavity resonance caused by the polarization dependence of the dipole
transition strengths and the cavity birefringence. The coupling
strength $2\lambda /\sqrt{N}$ is given by the two-photon (vacuum)
Rabi frequency between pump and cavity fields \cite{SBaumann2010}.

Residual driving of the cavity field due to finite-size effects of
the atomic cloud \cite{SBaumann2011} and residual scattering of pump
light off the cavity mirrors results in a finite coherent cavity
field amplitude already in the normal phase. The presence of this
field breaks the $\mathbb{Z}_2$ symmetry of the Dicke Hamiltonian
and is modeled by the replacement $\hat{J}_x \rightarrow \hat{J}_x +
\zeta$ in equation \eqref{eq:Dicke Hamiltonian}, where $\zeta$
(assumed to be real) determines the effective cavity drive
amplitude. Effectively, $\zeta$ can be interpreted as an atomic
population imbalance between even ($\cos(kx)\cos(kz) =1$) and odd
($\cos(kx)\cos(kz) =-1$) lattice sites at zero coupling strength
\cite{SBaumann2011}. Since in the experiment $\zeta$ is likely to
change in sign and amplitude between different experimental runs
(e.g.~due to drifts of the atomic cloud center with respect to the
pump-cavity mode structure), we include in the theoretical model an
ensemble average according to $\zeta \rightarrow \zeta\cos(\phi)$
where $\phi$ is taken to change randomly between 0 and $2\pi$.

\subsection{Semi-classical steady-state solution}

The self-consistent mean-field solution for $\alpha = \ev{\hat{a}}$,
$\beta = \ev{\hat{J}_-} \equiv \ev{\hat{J}_x-i \hat{J}_y}$ and $w =
\ev{\hat{J}_z}$ in the presence of a symmetry-breaking field ($\zeta
\neq 0$) is determined from the semi-classical equations of motion
of the Hamiltonian $\hat{H}+ 2 \hbar \lambda /\sqrt{N}
(\hat{a}+\hat{a}^\dag)\zeta$, including cavity decay at rate
$\kappa$:
\begin{eqnarray}
\dot{\alpha} &=& -(\kappa+i \omega) \alpha - i
\frac{\lambda}{\sqrt{N}}(\beta + \beta^* + 2 \zeta)\\
\dot{\beta} &=& -i \omega_0 \beta + 2 i
\frac{\lambda}{\sqrt{N}}(\alpha + \alpha^*)w\\
\dot{w} &=& i \frac{\lambda}{\sqrt{N}}(\alpha +
\alpha^*)(\beta-\beta^*)\,.
\end{eqnarray}
Using the conservation of $\mathbf{J}^2$, i.e.~$w^2 + |\beta|^2 =
N^2/4$, yields the steady-state equations
\begin{eqnarray}
\label{eq:steady-state meanfield solutions} \beta
&=&\frac{\lambda^2}{\lambda_\mathrm{cr}^2}(\beta +
\zeta)\sqrt{1-4\frac{\beta^2}{N^2}}\\
\alpha &=& \frac{2 \lambda}{i
\kappa-\omega}(\beta+\zeta)/\sqrt{N}\notag \,.
\end{eqnarray}
with critical coupling strength $\lambda\_{cr} =
\sqrt{\frac{(\kappa^2 + \omega^2) \omega_0}{4 \omega}}$, valid for
$\zeta=0$. In Fig. S1, the numerically obtained steady-state
solution $(\alpha, \beta, w)$ is plotted for typical experimental
parameters.

\begin{figure}[h]
\label{fig:SOM_SS}
\includegraphics[width=1\linewidth]{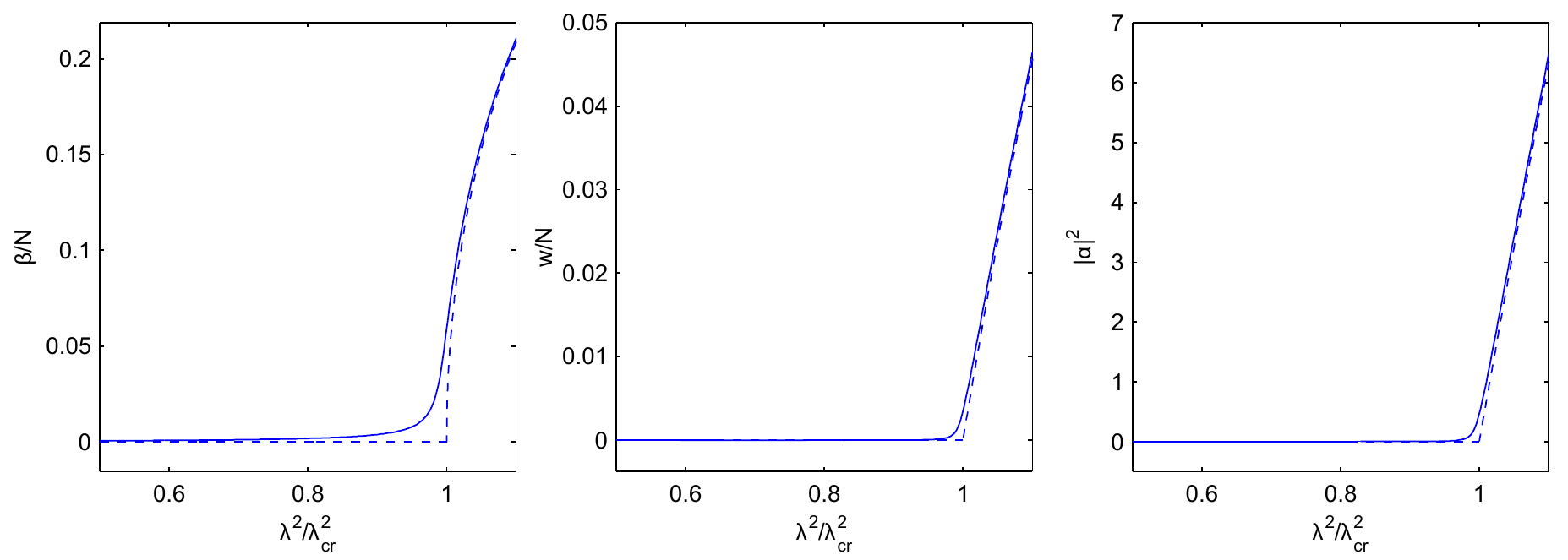}
\caption*{FIG. S1. Steady-state solution as a function of relative
interaction strength $\lambda^2/\lambda\_{cr}^2$ for our
experimental parameters and an order parameter offset of $\zeta =
65$ (solid curves). The dashed curves display for comparison the
case where no symmetry-breaking field is present, i.e. $\zeta = 0$
(only one branch of the bifurcation is shown). In the experiment,
the intracavity photon number $|\alpha|^2$ rises much faster for
$\lambda>\lambda_\mathrm{cr}$ as compared to the two-mode model
since the dispersive shift of the cavity frequency dynamically
increases with the emerging atomic density
modulation\cite{SBaumann2010}.}
\end{figure}

\subsection{Thermodynamic limit and quantum Langevin equations}

In the thermodynamic limit, $N \gg 1$, fluctuations of the system
around the semiclassical steady-state are small and can be treated
in a linearized approach \cite{SDimer2007}. This is based on the
Holstein-Primakoff transformation which maps the collective spin
degree of freedom $\hat{\mathbf{J}}$ to a bosonic mode with mode
operators $\hat{b}$ and $\hat{b}^\dag$, defined as
\cite{SLambert2004}
\begin{eqnarray}
\hat{J}_- &=& \hat{b}\sqrt{N-\hat{b}^\dag \hat{b}}\\
\hat{J}_z &=& \hat{b}^\dag \hat{b} - N/2\,.
\end{eqnarray}
Inserting this transformation into the Dicke Hamiltonian,
Eq.~\eqref{eq:Dicke Hamiltonian} yields to leading order in $N^{-1}$
the quadratic Hamiltonian
\begin{equation}\label{Dicke Hamiltonian_TL}
\hat{H}'/\hbar = \omega \delta\hat{a}^\dag \delta\hat{a} +
\tilde{\omega}_0 \delta\hat{b}^\dag \delta\hat{b} - \mu
(\delta\hat{b}+ \delta\hat{b}^\dag)^2 + \tilde{\lambda}
(\delta\hat{a}+\delta\hat{a}^\dag)(\delta\hat{b}+\delta\hat{b}^\dag)
\end{equation}
with fluctuation operators $\delta\hat{a} = \hat{a} - \alpha$ and
$\delta\hat{b} = \hat{b} - \beta/\sqrt{N}$ and renormalized
parameters
\begin{eqnarray}
\tilde{\omega}_0 &=& \omega_0 - 2\lambda
\frac{\mathrm{Re}(\alpha)\beta}{N^{3/2} \sqrt{1-\beta^2/N^2}}
\\
\tilde{\lambda} &=& \lambda\frac{1-2\beta^2/N^2}{\sqrt{1-\beta^2/N^2}}\\
\mu &=& -\lambda
\frac{\mathrm{Re}(\alpha)\beta}{N^{3/2}\sqrt{1-\beta^2/N^2}} \,.
\end{eqnarray}
For our experimental parameters and $\lambda \leq
\lambda_\mathrm{cr}$, the maximum relative deviation of the
renormalized parameters $\tilde{\omega}_0$ and $\tilde{\lambda}$
from their bare values $\omega_0$ and $\lambda$ are $0.3\%$ and
$0.7\%$. For simplicity, we therefore replace in the following
analysis ($\tilde{\lambda},\tilde{\omega}_0)$ by the bare values
$(\lambda,\omega_0)$. The squeezing term $\mu
(\delta\hat{b}+\delta\hat{b}^\dag)^2$ is also neglected in the
following as $\mu/\omega_0 < 2\times 10^{-3}$ for $\lambda \leq
\lambda_\mathrm{cr}$.

We model the driven-damped system dynamics with coupled quantum
Langevin equations of the form
\cite{SDimer2007,SNagy2011,SOztop2012}
\begin{eqnarray}\label{eq:Langevin}
\dot{\delta\hat{a}} &=& -i [\delta\hat{a}, \hat{H}'] -\kappa \delta\hat{a} +\sqrt{2 \kappa} \hat{a}\_{in}(t) \label{eq:lightLangevin}\\
\dot{\delta\hat{b}} &=& -i [\delta\hat{b}, H'] -\gamma \delta\hat{b}
+\sqrt{2 \gamma} \hat{b}\_{in}(t) \label{eq:atomicHeisenberg}.
\end{eqnarray}
In equation \eqref{eq:lightLangevin}, the bosonic operator
$\hat{a}\_{in}$ describes vacuum input fluctuations of the
surrounding electromagnetic field modes which are characterized by
the correlation functions $\langle
\hat{a}\_{in}(t)\hat{a}\_{in}^\dag(t')\rangle = \delta(t-t')$ and
$\langle \hat{a}\_{in}^\dag(t)\hat{a}\_{in}(t')\rangle = 0$.

In contrast to \cite{SDimer2007,SNagy2011,SOztop2012}, we
phenomenologically include an additional dissipation channel for the
collective atomic motional degree of freedom with effective damping
rate $\gamma$ accompanied by atomic input fluctuations described by
$\hat{b}_\mathrm{in}$. The physical origin for this dissipation
channel is attributed to collisional or cavity-mediated coupling of
excitations of the momentum mode $\psi_1$ to Bogoliubov modes with
wave vectors that are different from that of the pump and cavity
fields \cite{SGraham2000}. The collection of these modes provides a
heat bath at the condensate temperature $T$. For simplicity, we
assume this heat bath to behave Markovian. Correspondingly, we
require the correlation functions for the atomic input noise in
Fourier space to obey the relations
\begin{eqnarray}
\langle
\tilde{b}_\mathrm{in}(\nu)\tilde{b}_\mathrm{in}^\dag(\nu')\rangle =
2\gamma(1+n\_{th}(\nu))\delta(\nu-\nu')\\
\langle
\tilde{b}_\mathrm{in}^\dag(\nu)\tilde{b}_\mathrm{in}(\nu')\rangle =
2\gamma n\_{th}(\nu)\delta(\nu-\nu')
\end{eqnarray}
with thermal mode occupation number $n\_{th}(\nu) =
\frac{1}{\exp(\hbar \nu/k\_{B} T)-1}$. In the data analysis, the
damping rate $\gamma$ is taken as a free parameter which varies as a
function of the relative coupling constant
$\lambda/\lambda_\mathrm{cr}$. This corresponds to the expectation
that the density of states of the atomic heat bath, evaluated at the
softening excitation frequency $\omega_s$ of the coupled system,
changes when approaching the phase transition and vanishes at the
critical point.

Due to the coupling between atomic and optical fields, the lowest
lying excited states of the coupled BEC-cavity system acquire a
finite damping rate even for $\gamma = 0$ due to a finite admixture
of the cavity degree of freedom. However, the corresponding damping
rate $\frac{\lambda^2}{\lambda_\mathrm{cr}^2}\frac{\kappa
\omega_0^2}{\kappa^2+\omega^2} < 2\pi \times \unit[1]{Hz}$ is
negligible for our parameter regime, $\omega \gg \omega_0$, and does
not explain the observed damping of the cavity output correlation
function. Therefore we expect an additional atomic dissipation
channel to be present in the system.

We solve the coupled Langevin equations \eqref{eq:lightLangevin} and
\eqref{eq:atomicHeisenberg} in Fourier space according to the
conventions
\begin{align}
\tilde{O}(\nu) &= \frac{1}{\sqrt{2\pi}} \int_{-\infty}^\infty e^{i
\nu t} \hat{O}(t) \mathrm{d}t\\
\tilde{O}^\dag(-\nu) &= \frac{1}{\sqrt{2\pi}} \int_{-\infty}^\infty
e^{i \nu t} \hat{O}^\dag(t) \mathrm{d}t
\end{align}
for any given operator $\hat{O}$. We thus obtain the set of coupled
equations
\begin{equation}
\mathbf{M}(\nu)\begin{pmatrix} \delta\tilde{a}(\nu)\\
\delta\tilde{a}^\dag(-\nu)\\ \delta\tilde{b}(\nu)\\
\delta\tilde{b}^\dag(-\nu) \end{pmatrix} + \begin{pmatrix} \sqrt{2\kappa} \tilde{a}\_{in}(\nu)\\
\sqrt{2\kappa} \tilde{a}^\dag\_{in}(-\nu)\\ \sqrt{2\gamma}\tilde{b}\_{in}(\nu)\\
\sqrt{2\gamma}\tilde{b}\_{in}^\dag(-\nu) \end{pmatrix} = 0
\end{equation}
with the $4\times 4$ matrix
\begin{equation}
\mathbf{M}(\nu)=
\begin{pmatrix}
-i\nu +i \omega +\kappa & 0 & i \lambda & i \lambda\\
0 & -i\nu - i \omega + \kappa & -i \lambda & -i \lambda \\
i \lambda & i \lambda & -i\nu  +i \omega_0 +\gamma & 0 \\
-i\lambda & -i \lambda & 0 & -i\nu  - i \omega_0 + \gamma
\end{pmatrix}\,.\notag
\end{equation}

We note at this point that the cavity input noise operator
$\tilde{a}_\mathrm{in}(\nu)$ does not vanish for negative
frequencies $\nu$ in the frame rotating at the pump laser frequency
$\omega_\mathrm{p}$. This causes the cavity environment to act
e.g.~on the field quadrature $\hat{a}+\hat{a}^\dag$ effectively like
a thermal bath \cite{STorre2013}, and results in a distinct change
of the fluctuation spectrum of the driven-damped Dicke model with
respect to the ground state of the Dicke Hamiltonian.

The Langevin equations in Fourier space are solved by matrix
inversion. Denoting the matrix elements of $\mathbf{M}^{-1}$ by
$m_{ij}(\nu)$, we obtain e.g.
\begin{align}\label{solutions a Fourier}
\delta\tilde{a}(\nu) &= \sqrt{2\kappa} \big(m_{11}(\nu)
\tilde{a}\_{in}(\nu) + m_{12}(\nu)
\tilde{a}^\dag\_{in}(-\nu)\big)\notag\\
& + \sqrt{2\gamma} \big(m_{13}(\nu) \tilde{b}\_{in}(\nu) +
m_{14}(\nu) \tilde{b}\_{in}^\dag(-\nu)\big)\,,
\end{align}
with matrix elements
\begin{align}
D\times m_{11}(\nu) &= 2 i \lambda^2 \omega_0 -i (i\kappa+
\omega + \nu)((\gamma -i \nu)^2 +\omega_0^2)\\
D\times m_{12}(\nu) &= 2 i \lambda^2 \omega_0\\
D\times m_{13}(\nu) &= i\lambda(i\kappa + \omega + \nu)(i \gamma +
\nu + \omega_0)\\
D\times m_{14}(\nu) &= i\lambda (i\kappa + \omega + \nu)(i \gamma +
\nu - \omega_0)
\end{align}
and
\begin{equation}
D = \big((\gamma -i \nu)^2+\omega_0^2\big)\big((\kappa-i
\nu)^2+\omega^2\big) - 4 \lambda^2 \omega \omega_0
\end{equation}
denoting the determinant of $\mathbf{M}$. Analytical expressions for
the overlap integrals defined below can be obtained for our
experimental parameter regime, $\omega_0 \ll \omega$, by
approximating the determinant $D$ as
\begin{equation}\label{eq:determinant approximation}
D \backsimeq \big((\gamma -i
\nu)^2+\omega_s^2\big)(\kappa^2+\omega^2)\,,
\end{equation}
with the atom-like polariton frequency (soft mode frequency) defined
as $\omega_s = \omega_0 \sqrt{1-\lambda^2/\lambda_\mathrm{cr}^2}$
\cite{SMottl2012}. This approximation eliminates poles of the matrix
elements $m_{ij}(\nu)$ around $\nu = \pm \omega$ which do not
contribute to the critical behavior of the coupled system.

According to input-output theory, the cavity output field,
$\hat{a}\_{out}(t)$, is given by
\begin{equation}
\hat{a}\_{out}(t) = \sqrt{2\kappa} (\alpha + \delta\hat{a}(t)) -
\hat{a}\_{in}(t).
\end{equation}

\subsection{Correlation functions of the intracavity field}

\subsubsection{First-order correlation function} From
\eqref{solutions a Fourier} we obtain, using the correlation
functions of the input operators,
\begin{align}\label{G1 Fourier}
\bra \tilde{a}^\dag(\nu) \tilde{a}(\nu')\ket &= 2 \kappa
|m_{12}(\nu)|^2\delta(\nu-\nu')\\
&+2\gamma |m_{13}(\nu)|^2 n\_{th}(\nu)\delta(\nu-\nu')\notag\\
&+2\gamma |m_{14}(\nu)|^2
\big(1+n\_{th}(-\nu)\big)\delta(\nu-\nu')\notag\\
&+ 2 \pi |\alpha|^2\delta(\nu)\delta(\nu')\notag.
\end{align}
In the time domain, the first-order correlation function is given by
\begin{equation}
G^{(1)}(\tau = t-t') \equiv \bra \hat{a}^\dag(t) \hat{a}(t')\ket =
\frac{1}{2\pi}\int_{-\infty}^\infty \int_{-\infty}^\infty\bra
\tilde{a}^\dag(\nu) \tilde{a}(\nu')\ket e^{i\nu t-i\nu'
t'}\mathrm{d}\nu\mathrm{d}\nu'\,.
\end{equation}
Inserting Eq.~\eqref{G1 Fourier}, this reduces to
\begin{align}
G^{(1)}(\tau) &= \frac{\kappa}{\pi}A(\tau) +
\frac{\gamma}{\pi}\Big((1 + n\_{th}(\omega\_{s}))B(\tau) +
n\_{th}(\omega\_{s}) B^*(\tau))\Big) + |\alpha|^2\,,
\end{align}
where we defined overlap integrals
\begin{align}\label{eq:matrix element A}
A(\tau) &= \int_{-\infty}^\infty  |m_{12}(\nu)|^2 e^{i \nu
\tau}\mathrm{d}\nu\\
\label{eq:matrix element B} B(\tau) &= \int_{-\infty}^0
|m_{14}(\nu)|^2 e^{i \nu \tau}\mathrm{d}\nu =
\int_0^{\infty}|m_{13}(\nu)|^2 e^{-i \nu \tau}\mathrm{d}\nu
\end{align}
and made the approximation
\begin{align}
\int_{-\infty}^0 |m_{ij}(-\nu)|^2 n\_{th}(\nu) \mathrm{d}\nu \approx
 n\_{th}(\omega\_{s})\int_{-\infty}^0 |m_{ij}(\nu)|^2 \mathrm{d}\nu\,.
\end{align}
This simplification is justified by the fact that the matrix
elements $m_{ij}(\nu)$ peak at $\pm \omega_s$.

\subsubsection{Second-order correlation function}

To calculate the second-order correlation function $G^{(2)}(t-t') =
\langle \hat{a}^\dag(t)
\hat{a}^\dag(t')\hat{a}(t')\hat{a}(t)\rangle$\,, we use the relation
\cite{Snaraschewski1999}:
\begin{align}
G^{(2)}(t-t') &= |G^{(1)}(0)|^2 + |G^{(1)}(t-t')|^2 + |\bra
\hat{a}(t) \hat{a}(t')\ket|^2 - 2|\bra \hat{a}(t)\ket|^4\,,
\end{align}
and obtain
\begin{align}
G^{(2)}(\tau) =& |G^{(1)}(0)|^2 + |G^{(1)}(\tau)|^2 -2|\alpha|^4+\\
&\Big|\frac{\kappa}{\pi} C(\tau) +
\frac{\gamma}{\pi}\Big((1+n\_{th}(\omega\_{s})D(\tau) +
n\_{th}(\omega\_{s}) D^-(\tau))\Big)+\xi|\alpha|^2\Big|^2\notag
\end{align}
with $\xi = \frac{(\omega - i\kappa)^2}{\kappa^2+\omega^2}$,
$D^-(\tau) = \xi B^*(\tau)$ and matrix overlaps
\begin{align}
C(\tau)&= \int_{-\infty}^\infty m_{12}(\nu)m_{11}(-\nu)e^{i\nu
\tau}\mathrm{d}\nu\\
D(\tau)&=\int_{-\infty}^\infty m_{14}(\nu)m_{13}(-\nu)e^{i\nu
\tau}\mathrm{d}\nu\,.
\end{align}

Uncorrelated background light hitting our detector as well as
detector dark counts (corresponding effectively to a
mean-intracavity photon number of $n_B$) reduce the contrast of the
oscillations in the measured second-order correlation function. To
account for this, we define the normalized second-order correlation
function as $g^{(2)}(\tau) = \frac{G^{(2)}(\tau) + 2|G^{(1)}(0)|n_B
+ n_B^2}{|G^{(1)}(0) + n_B|^2}$, which is in very good agreement
with our data upon including the measured background count rate.

\subsection{Determination of density fluctuations from the cavity output signal}

As the lifetime of cavity photons $1/(2\kappa)$ is negligible
compared to the timescale of atomic motion (determined by
$\omega_0$), we can directly infer from the detected cavity output
field about the magnitude of atomic density fluctuations present in
the system. The adiabatic solution of equation
\eqref{eq:lightLangevin} reads
\begin{equation}
\delta\hat{a}(t) = \frac{\lambda (\delta\hat{b}+\delta
\hat{b}^\dag)}{i \kappa - \omega}+\sqrt{2 \kappa} \int_{-\infty}^t
e^{-(i \omega + \kappa) (t-t')} \hat{a}_\mathrm{in}(t') dt'.
\end{equation}
Correspondingly, the incoherent intracavity photon number $\bra
\delta\hat{a}^\dag \delta\hat{a}\ket$ reflects directly the variance
of the atomic order parameter (i.e. the variance of the checkerboard
density modulation):
\begin{equation}
\bra \hat{J}_x^2\ket = \frac{N}{4}\bra(\delta\hat{b}+\delta
\hat{b}^\dag)^2\ket = \frac{\lambda_\mathrm{cr}^2}{\lambda^2}
\frac{N\omega}{\omega_0}\bra \delta\hat{a}^\dag \delta\hat{a}\ket
\end{equation}
as $\bra \hat{a}_\mathrm{in}^\dag(t) \hat{a}_\mathrm{in}(t)\ket =
0$. After subtracting the coherent fraction $|\alpha|^2$ from the
total detected intracavity photon number, we thus are able to
extract the variance of atomic density fluctuations (Figure 5).

\subsection{Comparison between experimental data and theoretical model prediction}

From least-square fits of our theoretical model to the correlation
data (as shown exemplarily in Figure 3C), we deduce the decay rate
$\gamma$ as a function of $\lambda^2/\lambda_\mathrm{cr}^2$ (as
shown in Figure 4) and a maximum strength of the symmetry breaking
field given by $\zeta = 60(7)$. This corresponds (for $\lambda = 0$)
to a finite order parameter of $0.8$\textperthousand\, relative to
the maximum order parameter of $N/2 = 8 \times 10^4$. Since we
observe enhanced fluctuations for $\lambda^2 > 0.97
\lambda_\mathrm{cr}^2$ which are not explained by our model, we
deduce $\zeta$ from those correlation data where $\lambda^2 < 0.97
\lambda_\mathrm{cr}^2$.

We fit to the deduced $\gamma$ values the empirical function $f(x) =
c_1(1-x)^{c_2}\exp{(c_3 x)}+c_4(1-x)^{c_5}\exp{(c_6 x)}$ with $x =
\lambda^2/\lambda_\mathrm{cr}^2$ and free parameters $c_j$. The
obtained functional dependence $\gamma(\lambda)$ as well as the
deduced value of $\zeta$ enter the theory curves plotted in Figure
2.

In this analysis, the temperature of the atomic heat bath was set to
the expected temperature $T = \unit[100(20)]{nK}$ of the gas after
probing. This is justified by the fact that the thermal population
of the momentum mode $\psi_1$ becomes relevant only close to the
transition point due to the softening of the excitation spectrum.

\section{Error analysis}

The systematic uncertainty of the intracavity photon number
(estimated to $\pm 20\%$) is dominated by the uncertainty of the
detector quantum efficiency, the cavity mirror transmission and the
calibration of the detector background count rate. The statistical
uncertainty in the determination of $\lambda/\lambda_\mathrm{cr}$
originates from power fluctuations of the transverse pump beam and
residual dipole trap oscillations of the atomic cloud, which convert
into fluctuations of the critical pump power. The uncertainty of
$\zeta$ converts into an uncertainty of the deduced atomic density
fluctuations $\langle (\hat{b} + \hat{b}^\dag)^2\rangle$, as
indicated by the errorbars in Figure 5.

\end{widetext}

\bibliographystyle{plain}

\end{document}